\documentclass[crystals,article,submit,moreauthors,pdftex,10pt,a4paper]{mdpi} 
\usepackage{booktabs} 
\usepackage{multirow}
\usepackage{soul} 
\usepackage{microtype}
\usepackage{upgreek}
\graphicspath{{./Figures/}}
\firstpage{1} 
\articlenumber{x}
\doinum{10.3390/------}
\pubvolume{7}
\pubyear{2017}
\copyrightyear{2017}
\externaleditor{Academic Editor: Yuri Palyanov}
\history{Received: 8 April 2017 ; Accepted: 24 April 2017 ; Published: date}

\pdfoutput=1

\Title{Nanoscale Sensing Using Point Defects in 
Single-Crystal Diamond: Recent Progress on Nitrogen Vacancy Center-Based Sensors}

\Author{Ettore Bernardi, Richard Nelz, Selda Sonusen and Elke Neu *}
\AuthorNames{Ettore Bernardi, Richard Nelz, Selda Sonusen and Elke Neu}

\address[1]{%
Faculty for Natural Sciences and Technology, Physics Department, Saarland University, 66123 Saarbrücken, Germany; ettore.bernardi@physik.uni-saarland.de (E.B.); richard.nelz@uni-saarland.de (R.N.); selda.sonusen@physik.uni-saarland.de (S.S.)}

\corres{Correspondence: elkeneu@physik.uni-saarland.de; Tel.: +49-681-302-2739}

\abstract{Individual, luminescent point defects in solids, so-called color centers, are atomic-sized quantum systems enabling sensing and imaging with nanoscale spatial resolution. In this overview, we introduce nanoscale sensing based on individual nitrogen vacancy (NV) centers in diamond. We discuss two central challenges of the field: first, the creation of highly-coherent, shallow NV centers less than 10 nm below the surface of a single-crystal diamond; second, the fabrication of tip-like photonic nanostructures that enable efficient fluorescence collection and can be used for scanning probe imaging based on color centers with nanoscale resolution.  }


\begin{document}

\section{Introduction \label{sec:Introduction}}
Nanotechnology has led to many significant technological and scientific advances in recent years. For instance, two-dimensional or nanoscale materials such as carbon nanotubes or graphene are investigated for next generation electronics and photonics \cite{Dresselhaus2016}. Simultaneously, functionalized nanoparticles, e.g., for drug delivery, are promising to enhance various therapies \cite{Stefan2016}. Moreover, electronic systems like transistors are being miniaturized and controlled down to the single electron level \cite{Wagner2017}. 

Simultaneously to developing nanotechnology, a need for sensing techniques that work on the nanoscale has been arising to investigate nanoscale materials and to foster their further development. Quantities of interest are magnetic fields, often created as a result of electrical currents~\cite{Nowodzinski2015}, electric~fields~\cite{Dolde2014}, temperatures \cite{Kucsko2013}, pressure or crystal strain \cite{Momenzadeh2016}, as well as the presence of individual fluorescent markers, e.g., molecules \cite{Tisler2011}. For sensing with nanoscale spatial resolution, in general, the~sensor needs to fulfill several demanding prerequisites:
\begin{itemize}[leftmargin=1.8em,labelsep=4mm]
\item The sensor's size or active area has to be small compared to the structure under investigation. If this is not the case, spatial averaging over the detector area may mask information from the sample's nanostructure. Consequently, sensors approaching atomic dimensions (< 1 nm) are desirable for nanoscale sensing. 
\item The sensor's geometry must allow for close proximity in-between the investigated object and the sensor. In most cases, controlled proximity to the sample is ensured by manufacturing the sensor in a tip-like geometry and approaching it to the sample via a scanning probe mechanism. This~mechanism often keeps the force between the sample and the tip constant (for pioneering work see, e.g., \cite{Binnig1986,Martin1987,Ohnesorge1993}). Alternatively, the sensor can consist of a nanoparticle that is, e.g., inserted~into a cell for sensing \cite{Kucsko2013}.
\item The sensor needs to provide sufficient sensitivity to capture the weak signals that arise from nanoscopic or atomic objects. To illustrate this demanding point, the magnetic field of a single electron spin even at a distance of 50 nm amounts to only $\approx$ 9 nT \cite{Grinolds2013}. The field of magnetic dipoles decays with the distance $r$ from the dipole like $r^{-3}$ \cite{Rondin2014}. For the near-field energy transfer between two point-like dipoles, which is a valuable imaging resource, as well, even a $r^{-6}$ decay has to be considered \cite{Forster1948}. Thus, bringing the sensor and the sample in close proximity is not only mandatory for high resolution, but also enables detecting weak signals from nanoscale objects. 
 \end{itemize} 

Using individual, optically-active point defects in solids as sensors allows simultaneously fulfilling the prerequisites listed above. Moreover, such defects are versatile sensors for several quantities, including magnetic and electric fields and temperature.

Point defects alter the host crystal's periodic lattice only in one or a few neighboring lattice sites. Mostly, impurity atoms enter the crystal lattice and can form complexes with vacancies. Electrons, or more precisely the electronic wave-functions, localize at the defect within a few lattice constants and thus on atomic scales; typically within less than 1 nm \cite{Acosta2013}. This manifests also in the existence of such defects $\approx$ 1 nm below crystal surfaces \cite{OforiOkai2012}. At such distances, crystal surfaces strongly influence the defects and may cause instability \cite{Bradac2013}. Leading contenders for sensing are optically-active point defects in the wide-bandgap semiconductors diamond \cite{Zaitsev2001,Aharonovich2011,Rondin2014} and silicon carbide \cite{Kraus2014a}. 

\begin{figure}
\centering
\includegraphics[width=0.48\textwidth] {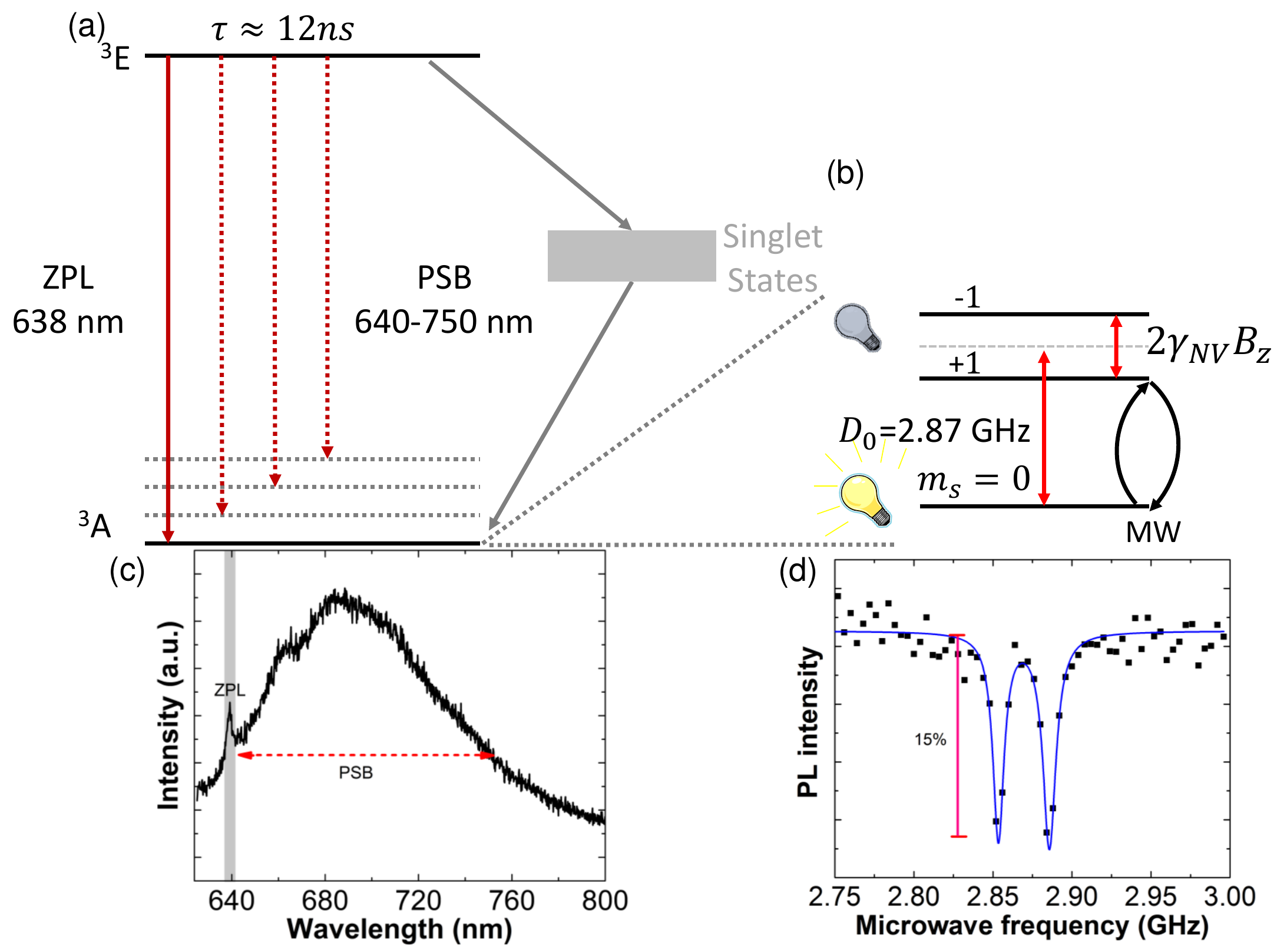}
\caption{\label{fig:NVbasics} Illustration of the basic properties of nitrogen vacancy (NV) centers in diamond. (\textbf{a}) Strongly-radiative, electric~dipole transitions between the NV excited (${}^3E$) and ground (${}^3A$) states create a photo-luminescence in the red and near-infrared spectral region shown in (\textbf{c}). For a detailed investigation of the NV level structure and dynamics, see, e.g., \cite{Manson2006a}. The excited state lifetime is \mbox{$\tau\, \approx$ 12--13 ns} in bulk diamond \cite{Collins1983}. The purely electronic transition at 638 nm (zero-phonon-line (ZPL)) is marked in the spectrum in (c) and features a width of about 1 nm at room temperature. Transitions to vibrationally-excited states create additional, broad phonon sidebands (PSB) as indicated in the schematics and in the spectrum. (\textbf{b}) Detailed description of the ground state spin levels. The~triplet state has three spin sub-levels: m$_s$=0~and m$_s$=$\pm$1. If no external field is applied, the levels with the projection spin quantum number 0 and $\pm$1 are split by 2.87 GHz due to spin-spin interactions. In the presence of a magnetic field, the Zeeman effect splits the +1 and $-$1 states by $2 \gamma_{NV} \textrm{B}_z$ as schematically shown in (b) and discernible from the measurement in (\textbf{d}). Note that only the magnetic field $\textrm{B}_z$ projected onto the NV's high symmetry axis (connecting line between vacancy and nitrogen, <111> direction) leads to a splitting \cite{Rondin2014}. If the NV center spin is in one of the $\pm$1 states, the probability is enhanced that the center undergoes an inter-system relaxation to the singlet levels (see (a)). As these levels have a lifetime that is more than one order of magnitude longer than for the triplet levels \cite{Robledo2011a}, the NV's photo-luminescence is reduced in the $\pm$1 states enabling the optical read-out of the NV spin state (optically-detected magnetic resonance (ODMR)). For sensing applications, transitions between the 0 state and +1 or $-$1 state are typically driven using circularly-polarized microwaves (MW) \cite{Alegre2007, Appel2015}. Using green laser light (532 nm), the NV center is initialized to its m$_s$=0 state via optical pumping within roughly 1 $\upmu$s \cite{Dreau2011}.}
\end{figure}

This review focuses on sensors based on diamond and especially one of the most prominent point defects in diamond, the nitrogen vacancy (NV) color center. This defect consists of a nitrogen atom replacing a carbon atom and a neighboring lattice vacancy \cite{Davies1976}. Figure \ref{fig:NVbasics} summarizes its basic properties. NV centers, in general, provide long-term photo-stable fluorescence. Their bright emission with high luminescence efficiency \cite{Radko2016} lies in the red and near-infrared spectral range and spans about 100 nm. NV color centers form emitting electric dipoles \cite{Alegre2007}. Their emission is bright enough to allow for a straightforward detection of individual, isolated centers in a confocal fluorescence microscope. Thus, NV centers have been investigated as solid-state sources of single photons \cite{Kurtsiefer2000}. Single photons in turn are a valuable resource for nanoscale sensing, e.g., in scanning near-field optical microscopy as a nanoscopic light source \cite{Kuehn2001}. Alternatively, the dipole of the NV center can interact with other dipoles and transfer energy via optical near fields (Förster resonance energy transfer (FRET) \cite{Tisler2011,Tisler2013}). Via~this process, NV centers can reveal the presence of other dipoles. 

In addition to their optical properties, NV centers provide highly-coherent, optically-readable electronic spins (see Figure\ \ref{fig:NVbasics}; first observation in\ \cite{Gruber1997}). The NV centers' long spin coherence times (T$_2^*$ or T$_2$) show that coherent superposition states of the 0 and $\pm$1 ground state spin levels retain their phase for a long time even at room temperature. Here, NV centers profit from the fact that the diamond lattice naturally features a low magnetic noise. Thus, it protects these superposition states from decoherence: the most abundant carbon isotope ${}^{12}C$ does not have a nuclear spin. Additionally, the concentration of paramagnetic ${}^{13}C$ isotopes (1.1\%) can be reduced by isotopically pure diamond synthesis \cite{Balasubramanian2009}. In such an isotopically-engineered diamond, T$_2$ can be as high as 1.8 ms. The NV center's T$_2$ is a valuable sensing resource: its decrease, for example, directly reveals the presence of magnetic molecules on the diamond surface \cite{Ermakova2013}. The use of spin coherence as a sensing resource opens up novel sensing schemes with potentially enhanced sensitivity compared to classical techniques. These novel approaches are typically summarized under the term of quantum sensing (for a recent review, see, e.g.,\ \cite{degen2016quantum}). However, sensing approaches that rely on the direct measurement of spin resonances (optically-detected magnetic resonance (ODMR); for an explanation, see Figure\ \ref{fig:NVbasics}) and their shift in magnetic fields (Figure\ \ref{fig:NVbasics}) also profit from highly-coherent NV centers: the NV's sensitivity $\eta$ to static magnetic fields is given by \cite{Rondin2014}:
\begin{equation}
\eta \approx \frac{h}{g \mu_B} \frac{\Delta \nu}{\sqrt{I_0} C}\,\,,
\label{eq:sens_DC}
\end{equation}
where $\Delta \nu$ is the ODMR linewidth and $I_0$ is the detected photon rate from the NV center. $C$ is the fluorescence contrast in ODMR. The latter is an intrinsic property of NV centers and can hardly be modified as it is determined by the internal dynamics. $\Delta \nu$ is fundamentally limited by the inverse of the coherence time T$_2^*$ and thus connects the sensitivity to the coherence properties.

Note that Equation\ (\ref{eq:sens_DC}) refers to non-resonant spin read-out, typically using green laser light to excite the NV center. This read-out scheme is feasible at room temperature and thus advantageous for sensing. At cryogenic temperature, resonantly-addressing spin-selective transitions within the ZPL enables single-shot read-out of the electronic spin state of NV centers \cite{Robledo2011}.

The carbon lattice of diamond itself provides only a weak source of decoherence. However, paramagnetic impurities (e.g., nitrogen in its substitutional form) and spins on the diamond surface can significantly reduce the coherence time of NV centers \cite{Luan2015}. Reduced coherence times render NV centers less sensitive for magnetic fields, as discernible from Equation\ (\ref{eq:sens_DC}). Consequently, it is vital to control this loss of coherence to use NV centers as highly-sensitive sensors for fields outside the diamond crystal. 

In essence, realizing an optimal NV sensor narrows to three main aspects:
\begin{itemize}[leftmargin=1.8em,labelsep=4mm]
\item It is mandatory to reliably create stable NV centers with a controlled density buried less than $\approx$ 10~nm below diamond surfaces ({shallow NV centers}). These NV centers need to retain spin coherence for optimal sensitivity. 
\item ODMR and optical sensing schemes demand efficient collection of fluorescence light and high photon rates from single centers. This in turn demands the incorporation of color centers into nanophotonic structures.
\item To realize a sensor that probes the sample surface and to realize controlled positioning of the NV sensor requires realizing a tip-like sensor and scanning probe sensing schemes. 
\end{itemize}
This review is structured according to these main aspects. In Section \ref{sec:NVcreation}, we summarize recent progress in creating and optimizing shallow NV centers. In Section \ref{sec:nanostructure}, we turn to the photonic nanostructures for sensing and their fabrication. In Section \ref{sec:advances}, we illustrate scanning probe-based approaches, as well as recent advances in NV-based sensing.

\section{Shallow NV Centers for Sensing \label{sec:NVcreation}} \vspace{-12pt}
\subsection{Creation Methods and Creation Yield \label{subsec:NVcreation}}
\begin{figure}[H]
\centering
\includegraphics[width=15cm]{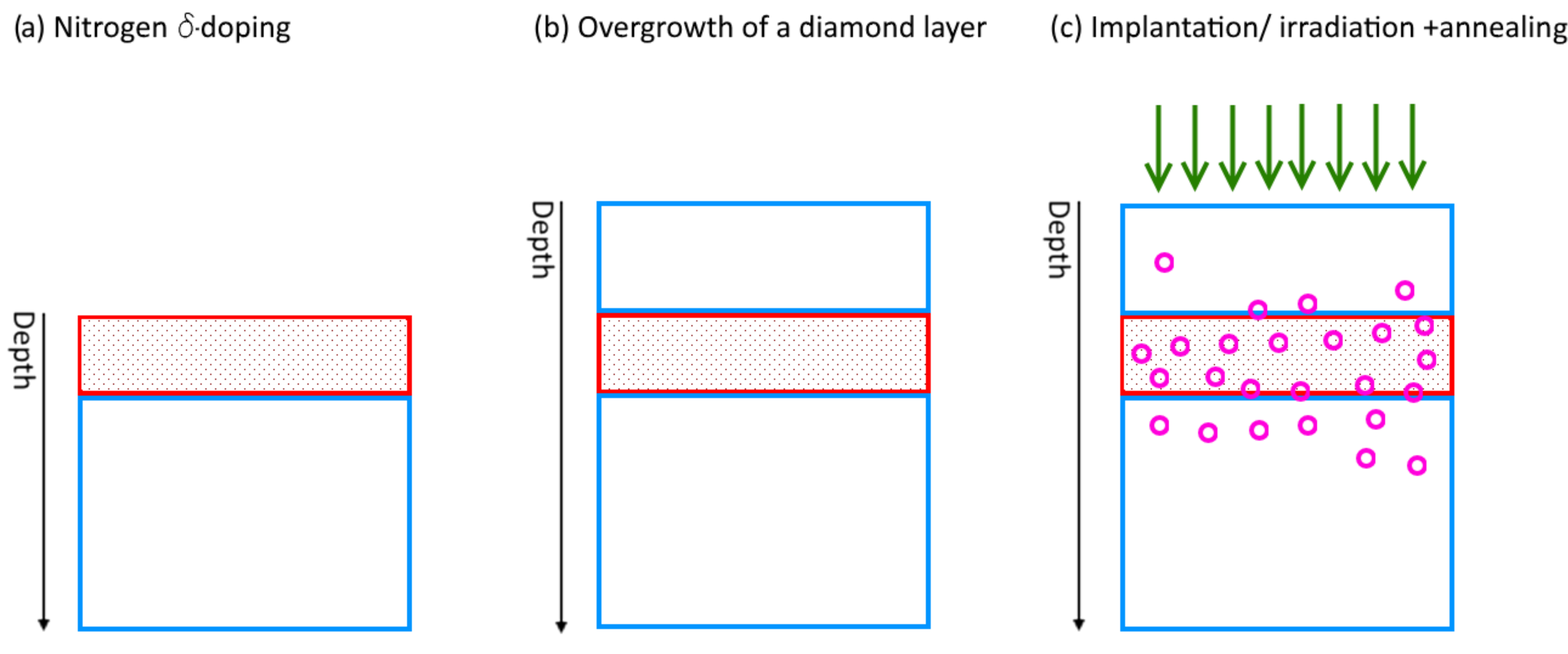}
\caption{Schematic representation of the $\delta$-doping method. (\textbf{a}) Creation of an N-doped, several nm-thick layer, the $\delta$-doped layer (red, spotted rectangle). (\textbf{b}) Overgrowth with an undoped, epitaxial diamond layer (blue rectangle). (\textbf{c}) Implantation/irradiation and annealing. Pink circles indicate lattice vacancies close to the $\delta$-doped layer.}
\label{deltadoping}
\end{figure}

NVs with a controlled distance to diamond surfaces have been created using two approaches, namely~ion implantation and $\delta$-doping. We discuss both in the following: 
\begin{itemize}[leftmargin=1.8em,labelsep=4mm]
\item {Ion implantation:} This approach relies on the commercial availability of chemical vapor deposition (CVD) diamond with low nitrogen (N) content (< 5 ppb substitutional N) and an almost negligible density of in situ-created, native NV centers (\cite{Ryan2010}, element six, electronic grade diamond). This~diamond is irradiated with N ions at energies typically between 4 and 8 keV. These~implantation energies correspond to mean implantation depths between $\approx$ 7 nm and \mbox{$\approx$ 12 nm,} as calculated by Monte Carlo simulations \cite{Ziegler2010}. Coherence times T$_2$ of NVs created deep (> 50~nm) inside this material by N implantation reach T$_2$ $\approx$ 200 $\upmu$s comparable to in situ-created NV centers~\cite{Wangjunfeng2016}. Such T$_2$ times indicate the high purity of the material, whereas in diamonds with 100 ppm N, T$_2$ generally reduces to about 1 $\upmu$s \cite{Rondin2014}. After the implantation, high temperature annealing repairs crystal damage, mobilizes vacancies via diffusion and forms NV complexes via N impurities capturing vacancies. In contrast, impurities like, for example, N are not expected to diffuse at these temperatures \cite{OforiOkai2012}. Two approaches are reported: annealing in vacuum at pressures below $5\times10^{-7}$ mbar \cite{Appel2016,Wangjunfeng2016,OforiOkai2012}, where a high vacuum is needed to avoid etching of the surface \cite{Antonov2014}. Alternatively, annealing in forming gas (4\% H$_2$ in Ar) is used (e.g., \cite{Orwa2011, Santori2009}). The conversion efficiency from implanted N to NV (mostly NV$^-$) is called yield. It amounts to typically only $1\%$ for an implantation energy of 5 keV (depth $\approx$ 8 nm) \cite{Pezzagna2010, Appel2016}. For micrometer-deep implanted NVs ($E=18$ MeV), the yield increases to $45\%$ \cite{Pezzagna2010}. For shallow NVs, vacancies are partially captured by the diamond surface \cite{Antonov2014}. Furthermore, higher energy implantation increases the number of vacancies produced \cite{Pezzagna2010}. Very recent studies, however, indicate that increasing the number of vacancies by co-implanting, e.g., carbon, does not increase the NV yield for shallow implantation \cite{DeOliveira2016b}.

\item The $\delta$-doping method for creating shallow NVs consists of the following steps; see Figure\ \ref{deltadoping}:
   \begin{itemize}[leftmargin=1.8em,labelsep=4mm]
   \item An N-doped, several nm-thick layer, the $\delta$-doped layer \cite{Ohno2012,Osterkamp2015}, is created in situ by the controlled introduction of N$_{2}$ gas during slow, plasma-enhanced CVD growth of single-crystal diamond (growth rate $\approx$ 0.1 nm/min; Figure\ \ref{deltadoping}a). Changing the N$_{2}$ flow tunes the resulting N densities. In order to decrease the magnetic noise, diamond growth can be performed using isotopically-purified $^{12}$CH$_{4}$ as the carbon source \cite{Ohno2012,Ohno2014,McLellan2016}. Very recently, overgrowth of a nitrogen-terminated diamond surface has been employed for $\delta$-doping \cite{Chandran2016}.

   \item Vacancies are created ex situ by implanting helium ions \cite{DeOliveira2016a,Kleinsasser2016}, carbon ions \cite{Ohno2014} or irradiating with electrons \cite{Ohno2012,McLellan2016}. A subsequent annealing at high temperature causes vacancy diffusion and creates NV centers (Figure\ \ref{deltadoping}c). Varying the thickness of an undoped diamond layer, overgrown before the implantation step, allows controlling the depth of the NV layer (Figure~\ref{deltadoping}b). The vacancy profile depends on the species used: for electron irradiation, it is flat and extends throughout the grown film and substrate \cite{Ohno2014}. In contrast, implanted carbon ions are localized in the lattice. A shallow layer ($\approx$ 5 nm) of implanted $^{12}$C ions has been used as the source of vacancies for a deeper ($\approx$ 50 nm) $\delta$-doped layer \cite{Ohno2014}, taking advantage of annealing-induced diffusion. With this method, it is possible to increase the NV density in the $\delta$-doped layer by increasing the $^{12}$C dose without activating NV centers in the substrate. An alternative method, applied to low energy helium implantation, consists of growing a thicker N-doped layer ($d=18$ nm) and adding a last etching step to create NV centers near the surface \cite{DeOliveira2016a}.
   \end{itemize}
         
NV yields reached values of $15\%$ \cite{DeOliveira2016a} and $50\%$ \cite{McLellan2016}, however, remaining quite low ($1.9\%$) in~\cite{Ohno2014}.
    As discussed below, often, $\delta$-doping enhances the coherence times of shallow NVs compared to N implantation. Moreover, it potentially localizes NV centers in a more defined~depth.
\end{itemize}

\subsection{Photochromism, Quantum Efficiencies and Coherence Time}
NV centers exist in two luminescent charge states: negative (NV$^-$) and neutral (NV$^0$) \cite{Mita1996}; only~NV$^-$ exhibits ODMR. To create an NV$^-$, an electron has to be captured by the NV center. The NV$^-$ population never exceeds $75\%$--$80\%$ (excitation wavelength 450--610 nm) \cite{Aslam2013,Chen2013,Beha2012}, due to photo-induced ionization \cite{Beha2012}. In the band from 450--575 nm, NV$^-$ and NV$^0$ absorb light. Consequently, the loop of NV$^-$ excitation-ionization and NV$^0$ excitation-recombination is closed, and the NV cycles between both charge states. The NV$^-$ steady state population is maximized for optical excitation in the band 510--540~nm \cite{Aslam2013}. Besides this photochromism, the NV charge state is directly influenced by the Fermi level position in diamond. For shallow NVs, the Fermi level position closely relates to the electron affinity of the surface (see Section \ref{subsec:stab_impr}).

An important characteristic for light-emitting systems is their quantum efficiency (QE). It is defined as $QE=\frac{k_r}{(k_r+k_{nr})}$, where $k_r$, $k_{nr}$ are the radiative and non-radiative decay rate of the NV center, respectively. A higher QE, in general, leads to brighter emission from the centers. For shallow NVs, QE can be as high as $70\%$ ($82\%$) for a depth of 4.1 nm (8 nm), respectively \cite{Radko2016}. The work in \cite{Radko2016} estimates QE > 96\% in bulk diamond. This result implies decreased QE for shallow NVs, which was attributed to non-radiative decays induced by surface strain. Generally, careful control of strain and crystalline quality seems necessary to obtain high QE: QE in the range of only $25\%$--$60\%$ was found in ion-damaged diamond (H-implantation, fluence $F=1\times10^{15}$ cm$^{-2}$ corresponding to an estimated induced vacancy density $d=60$ ppm, \cite{Monticone2013}). NV centers in 25-nm nanodiamonds even may show QE~<~20\% \cite{Mohtashami2013}.

Generally, shallowly-implanted NV centers show reduced T$_2$ together with broadened ODMR lines \cite{OforiOkai2012,Romach2015, Appel2016, Maletinsky2012}. Table \ref{tabNV} illustrates this reduction in comparison to deep, native centers. Degraded~coherence properties are related to noise created by the proximity to the surface. The work in~\cite{Romach2015} presents a spectroscopic analysis of the noise (N dose 10$^8$ cm$^{-2}$). High-frequency noise arises due to surface-modified phonons and low frequency noise due to the surface electronic spin bath. For~ion doses of 10$^{11}$ cm$^{-2}$, noise produced by implanted defects close to NV centers adds to surface-related noise. Indeed, N bombardment introduces N impurities and vacancy complexes that act as paramagnetic centers, degrading the coherence time of NV centers \cite{Yamamoto2013,DeOliveira2017}.

\begin{table}[H]
\caption{ Typical value of T$_2$ for shallowly-implanted and native, bulk NV from \cite{OforiOkai2012}. Note that for centers at a depth of around 50 nm, the coherence time of deep centers is restored \cite{Wangjunfeng2016}.}
\centering
\begin{tabular}{cccc}
\toprule
\textbf{}	& \textbf{Photocounts}	& \boldmath{T$_2$ ($\upmu$s)} & \textbf{Linewidth (MHz)}\\
\midrule
Very Shallow NV$^-$ (2.1 nm)		  & $5\times10^4$			     & $12.2\pm0.6$        & 2\\
Shallow NV$^-$	(7.7 nm)      	& $4\times10^5$			     & $40.4\pm0.8$        & 1.2\\
Native NV$^-$	(6 $\upmu$m)      	& $4\times10^5$			     & $128\pm10$        & 1\\
\bottomrule
\end{tabular}
\label{tabNV}
\end{table}

\subsection{Methods to Improve Stability and Photoluminescence \label{subsec:stab_impr}}
\begin{figure}[H]
\centering
\includegraphics[width=15 cm]{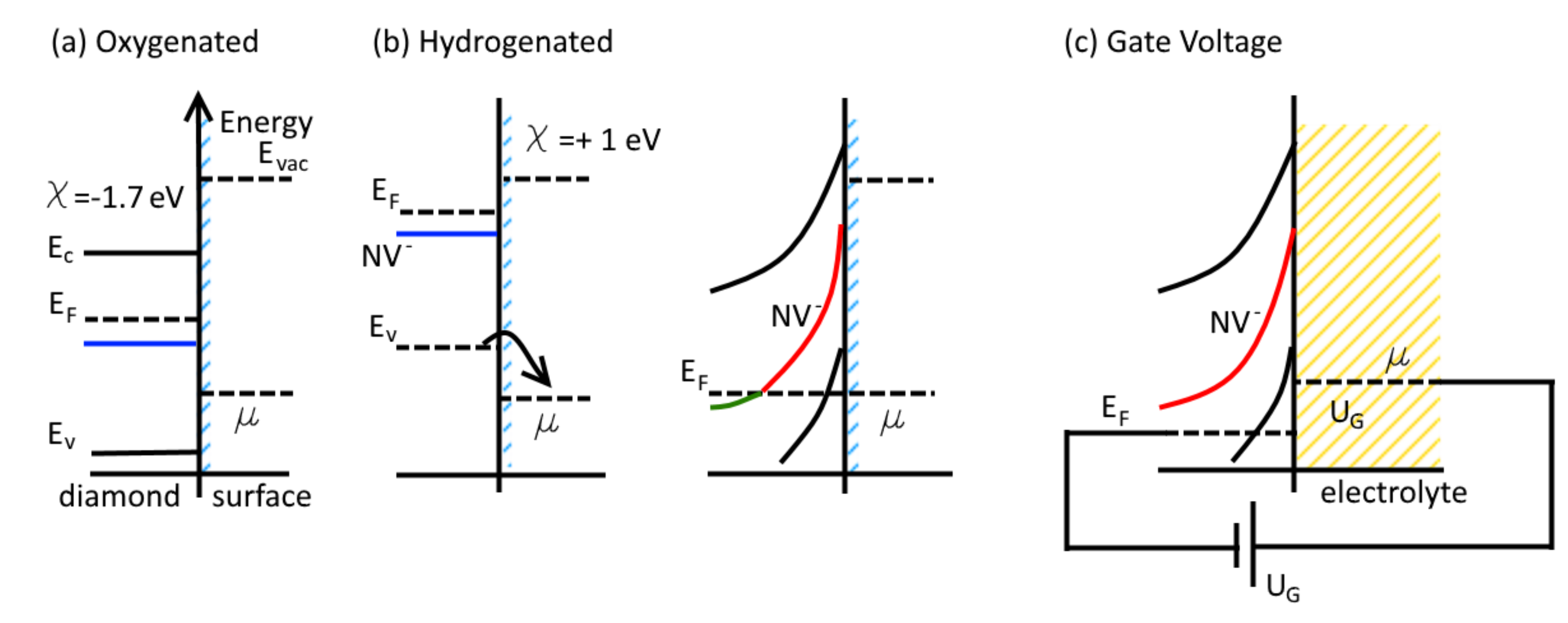}
\caption{Energy band schematic of diamond energy levels for: (\textbf{a}) oxygenated surface; (\textbf{b})~hydrogenated surface. (\textbf{c}) Applying a gate voltage in the presence of an electrolyte.}
\label{energylevels}
\end{figure}

The work in \cite{Santori2009} shows that the NV$^-$ population is decreased at depths up to 200 nm compared to bulk. This effect is attributed to the presence of an electronic depletion layer at the etched diamond surface. In this depletion region, N donors cannot donate an electron to the NV center because they are ionized. Stabilizing the NV$^-$ state close to the surface involves controlling the Fermi level in diamond~\cite{Fu2010,Hauf2011,Doi2014,Doi2016}.
The Fermi level, and consequently the NV$^-$ population, can be controlled via chemical functionalization of the surface \cite{Fu2010,Hauf2011,Shanley2014,Cui2013} (see Figure\ \ref{energylevels}a,b) by applying an electrolyte gate voltage \cite{Grotz2012} (see Figure\ \ref{energylevels}c) using in-plane gate nanostructures \cite{Hauf2014} or via doping the diamond \cite{Doi2016}.

First, we consider the decrease in the NV$^-$ population of shallow NV centers close to a hydrogen (H)-terminated surface compared to an oxygenated surface \cite{Hauf2011,Fu2010}. For H-terminated diamond, the conduction and valence bands shift upwards due to the negative electron affinity of the surface. The energy of the valence band maximum E$_V$ is higher than the chemical potential $\mu$ of the electronic states created by adsorbed water. Electrons migrate to the adsorbate layer. In equilibrium, the diamond bands bend upwards; a 2D hole gas is created; and the NV$^-$ charge state is depleted \cite{Hauf2011}. Thus, H-termination is not useful to stabilize NV$^-$.

In contrast, fluorine (F)-terminated surfaces will lead to a downward bending of the diamond bands due to the large electron affinity of F \cite{Rietwyk2013}. In this case, the NV$^-$ population increases compared to H-terminated diamond \cite{Shanley2014} and also to oxygenated diamond \cite{Cui2013}. However, it should be noted that so far, coherence measurements of shallow NV centers close to F-terminated surfaces are missing.

For H-terminated diamond, electrolytic gate electrodes can directly control the Fermi level \cite{Grotz2012} (see Figure\ \ref{energylevels}c). An increase in the NV$^-$ ensemble population was observed for high implantation dose and positive gate voltage. Unfortunately, no single NV charge switching was observed, due to the high gate voltage needed for charge-state conversion at low implantation dose. On the other hand, this result was achieved in\ \cite{Hauf2014} by applying an electric field using in-plane nanostructures. The~gate-voltage was applied to a structure formed by H-terminated areas and an O-terminated line, resulting in an offset between the Fermi level in the H-terminated areas on both sides of the line. In~this way, switching~the charge state of a single NV center from NV$^0$ to NV$^-$ was established.

Deterministic electrical control of the charge state of a single NV center has been achieved using a p-i-n diode \cite{Doi2014}. The NV center is positioned in the intrinsic region of the diode, and a current of holes is induced from the p-region. However, this technique converts NV centers to NV$^0$. Using~in plane Al-Schottky diodes, based on H-terminated diamond, charge state switching of single NV centers (NV$^+$ to NV$^0$ to NV$^-$) has been obtained \cite{Schreyvogel2015}. Implementing diode structures in scanning probe sensing seems, however, highly challenging.

Another way to change the Fermi level of diamond is by boron or phosphorus (P) doping \cite{Doi2016,Groot-Berning2014}. In particular, P impurities donate an electron to NV$^0$, due to the fact that the P activation energy is low ($E_A=0.57$ eV) compared to NV acceptor states ($E_{NV}=2.58$ eV). In this way, a~five-fold increase in luminescence and a pure NV$^-$ state was observed for a single NV \cite{Doi2016} and a P doping \mbox{$P=5\times10^{16}$~cm$^{-3}$}. It should be noted that this single NV had a short coherence time T$_2$~=~(19.77~$\pm$~0.27)~$\upmu$s, and the depth of the NV center in this case was not reported.

\subsection{Methods to Improve T$_2$}
\begin{figure}[H]
\centering
\includegraphics[width=\textwidth]{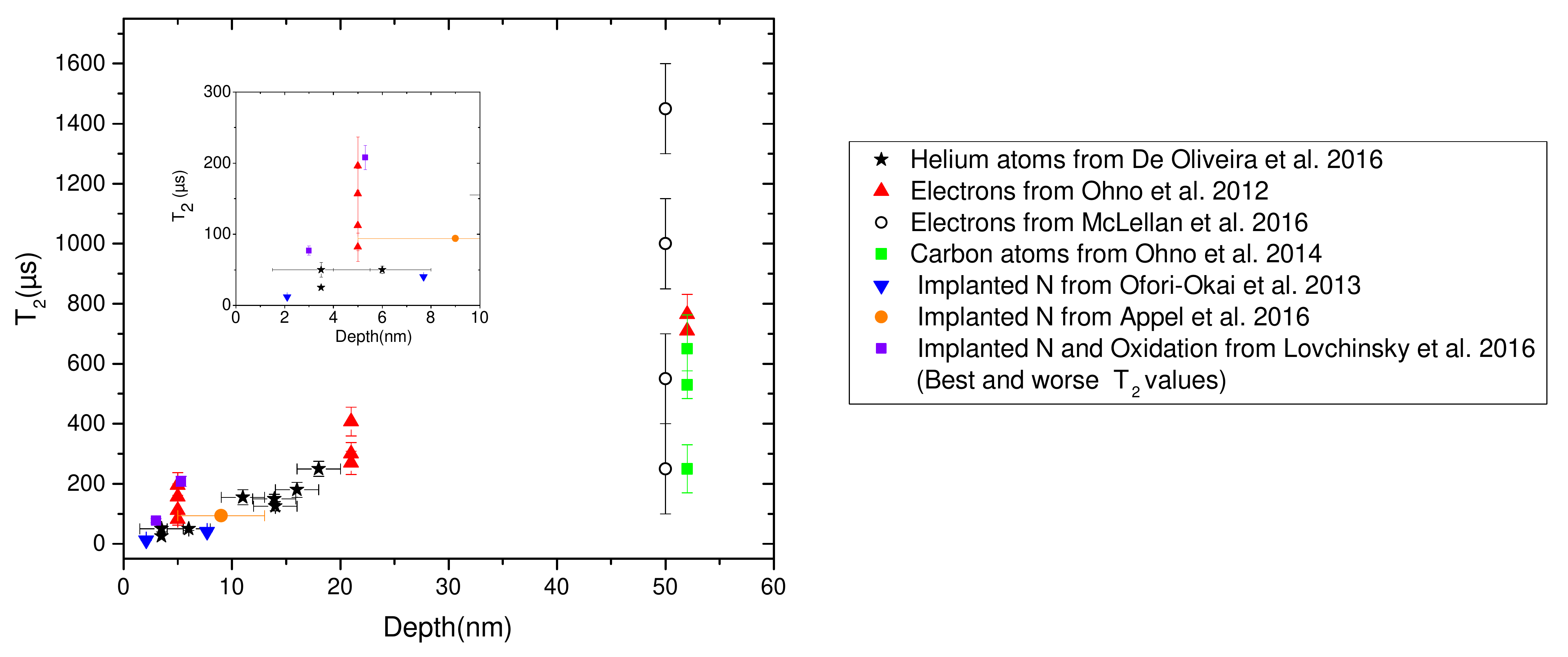}
\caption{Coherence time T$_2$ for NV centers created via $\delta$-doping methods and implantation with different inert species. For comparison, T$_2$ of centers created via N implantation are given. The inset highlights T$_2$ for an NV center less than 10 nm below the surface, which are especially important for sensing applications. Note the spread of T$_2$ in the different works. }
\label{T2fig}
\end{figure}

In this section, we discuss how N $\delta$-doping \cite{Ohno2012,Chandran2016} and ion implantation \cite{Huang2013,DeOliveira2016a,Naydenov2010} or electron irradiation \cite{McLellan2016} can lead to improved spin coherence. Figure \ref{T2fig} summarizes recently obtained T$_2$ values for different methods and works. 
As discussed above, implanted defects are sources of noise.
Implanted defects can be of two types:

\begin{itemize}[leftmargin=*,labelsep=4mm]

\item Vacancy complexes \cite{DeOliveira2017};

\item Ion impurities and crystal defects \cite{Wang2013}.

\end{itemize}

The formation of vacancy complexes during annealing is inhibited by Coulomb repulsion, if~their charge state is changed from neutral to positive \cite{DeOliveira2017}. Vacancy charging is accomplished in a p-i junction by implanting N in the junction's space charge layer. A two-fold increase in yield and ten-fold increase in T$_2$ for very shallow NVs has been reported (T$_2$ = 30 $\upmu$s, NV depth d = 3 nm).

Noise from implantation-induced impurities and crystal defects can be reduced using $\delta$-doping and ex situ creation of vacancies (see Section \ref{subsec:NVcreation}). To be used in scanning probe sensing, NV centers should have a depth below $\approx$ 10 nm. As mentioned before, this can be achieved by tuning the dimension of the cap layer \cite{Ohno2012} or adding a last etching step \cite{DeOliveira2016a}. In this depth range, the maximum reached T$_2$ is 200 $\upmu$s obtained when irradiating an isotopically-purified sample with electrons \cite{Ohno2012}. In this case, however, the NV density was too low for applications requiring the fabrication of single-crystal scanning probes, as discussed below.

We remark that NV areal densities around $D_{NV}=3 \times 10^9$ cm$^{-2}$ are typically necessary for single crystal scanning probes: this corresponds to one NV center in a circular area with a diameter of 200~nm and thus to a single NV in typical nanostructures used for sensing (see Section \ref{sec:nanostructure}). In~\cite{Ohno2012}, the~$\delta$-doped layer has an N concentration of $C_N=3\times10^{16}$ cm$^{-3}$, considering~a layer thickness of 2 nm; this results in an areal density of N of $6\times10^9$ cm$^{-2}$, so in order to have an acceptable NV areal density, one should have a very high yield, around $50\%$. A similar concentration of N \mbox{($C_N=0.8\pm0.6\times10^{16}$ cm$^{-3}$)} is reported in \cite{Ohno2014}, leading to a low NV area density \mbox{($D_{NV}=10^6$ cm$^{-2}$)}. A slightly too low NV density is reported in \cite{McLellan2016} $D_{NV}=1.4\times10^9$ cm$^{-2}$ and in~\cite{DeOliveira2016a} \mbox{$D_{NV}=1\times10^8$ cm$^{-2}$}. It is worth noticing that an N volume concentration of $C_N=1.8\times10^{20}$ cm$^{-3}$ is reported in \cite{Chandran2016}; this would result in an NV density of $D_{NV}=4\times10^{12}$ cm$^{-2}$, considering a thickness layer of 2 nm and a yield of $10\%$. In contrast, the N implantation method described in\ \cite{DeOliveira2017} seems more promising from this point of view: tuning the fluence of implanted N from $10^{10} $cm$^{-2}$--$10^{12}$ cm$^{-2}$ (E = 5 keV) allows accessing the optimal densities of shallow NV centers.

We underline that till now, $\delta$-doping has not been applied in the manufacturing of diamond probes for nanosensing. This is probably due to the fact that $\delta$-doping involves the realization of sophisticated CVD methods. In contrast, creating single NVs by ion implantation can be achieved using commercial diamond material and implantation facilities.

Finally, we remark that surface oxidation has been used recently to improve T$_2$ \cite{Lovchinsky2016}. Wet oxidative chemistry and sample annealing at 465 $^\circ$C in a dry oxygen atmosphere increased T$_2$ by an order of magnitude (max.\ T$_2=208$ $\upmu$s for a single NV, depth $(5.3\pm 0.1)$ nm, T$_2>100$ $\upmu$s observed for 6 NVs). Recent work \cite{Yamano2017}, however, was not able to reproduce these results for shallower NVs: in~\cite{Yamano2017}, a~similar oxidation procedure resulted in T$_2$ $\leq6$ $\upmu$s for calculated depths of \mbox{$(2.6\pm 1.1)$ nm}.

\section{Nanostructures for Photonics and Scanning Probe Operation \label{sec:nanostructure}}

NV centers close to the surface of bulk, single-crystal diamond are, on the one hand, advantageous as they reside in high-quality, high-purity, potentially low-stress diamond material mostly synthesized by the CVD method \cite{Balmer2009}. On the other hand, these NV centers do not fulfill all prerequisites for nanoscale sensing: first, positioning an NV created in a macroscopic diamond crystal within nanometer distance from a sample is hardly feasible. In principle, attaching the `sample', e.g., the substance under investigation, to a scanning probe tip while keeping the NV center stationary can circumvent this issue. However, this limits the technique to microscopic samples, e.g., specific molecules or ions (e.g.,~\cite{Pelliccione2014}) for which attachment to a nano-sized tip is feasible. 

\begin{figure} [H]
\centering
\includegraphics[width=0.8\textwidth]{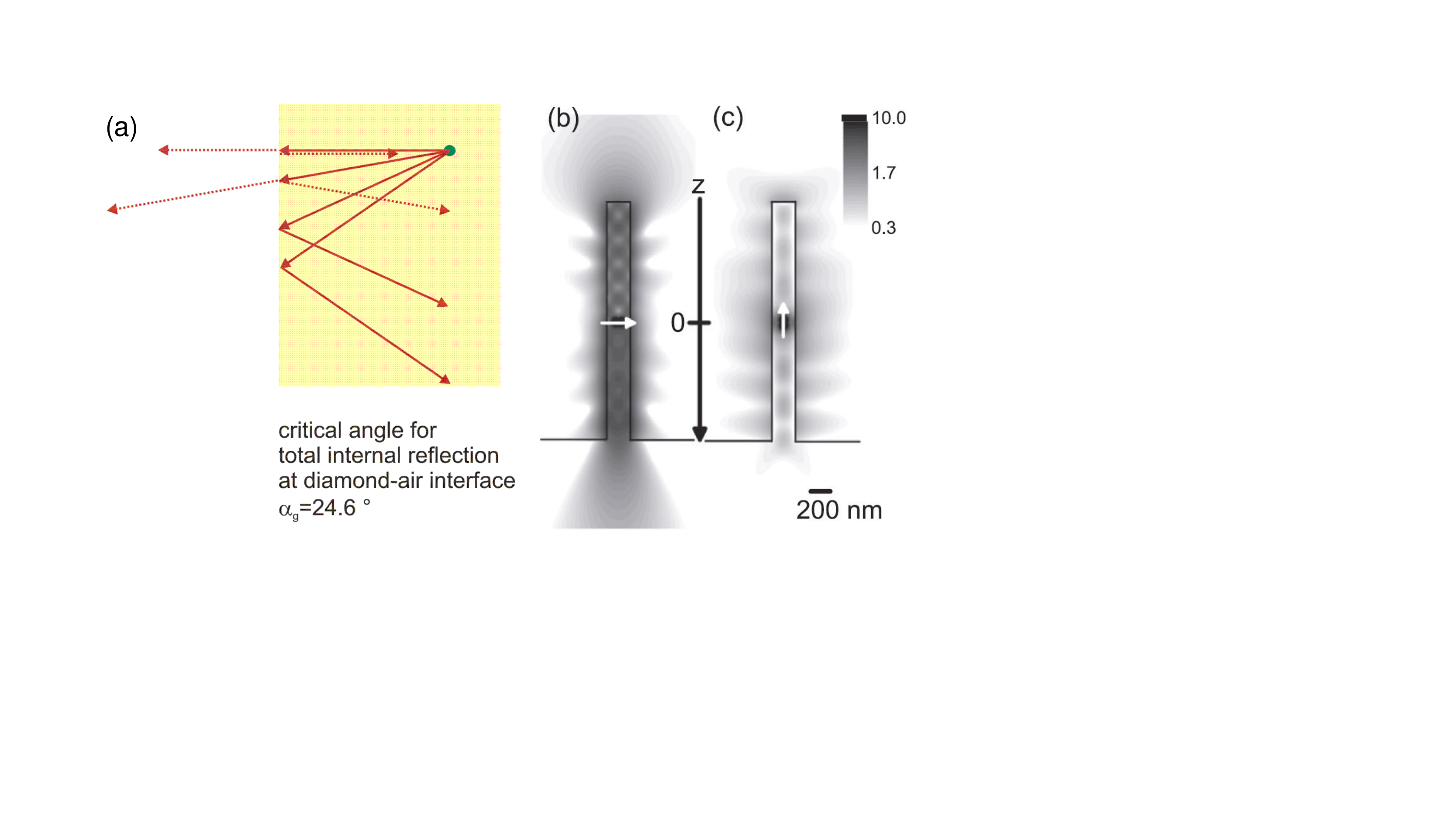}
\caption{\label{fig:tirphotonicstructures} (\textbf{a}) Illustration of total internal reflection: light incident under low angles cannot leave the diamond; for normal incidence, roughly 17\% of the incident power is reflected (Fresnel reflection). (\textbf{b},\textbf{c}) These images show the magnitude of the electric field (simulated for a light wavelength of \mbox{$\lambda$ = 700 nm}) inside diamond nanopillars. Note that darker areas correspond to higher fields, as indicated in the color bar. The field is being channeled in the direction of the pillar axis, witnessing wave-guiding via the photonic modes of the pillars. Reprinted from\ \cite{Neu2014} under CC-BY
 3.0. The arrow within the structures indicates the dipole orientation. For the optimal dipole orientation, where the dipole axis is perpendicular to the pillar axis, collection efficiencies up to 40\% can be reached (collection into NA
 0.8 from the air side). }\vspace{-6pt}
\end{figure}

Additionally, even for an optimal dipole orientation, a typical air microscope objective (NA~0.8~\cite{Neu2014}) collects only about 5\% of the fluorescence from an NV center in bulk diamond: as diamond exhibits a high refractive index of 2.4 for visible light \cite{Zaitsev2001}, only light incident at angles < 24.6$^\circ$ (partially) leaves the diamond as illustrated in Figure\ \ref{fig:tirphotonicstructures}a. Other light rays are lost due to total internal reflection. Additionally, the dipolar emission from color centers is typically directed towards the optically-dense medium. Thus, in our case, emission is directed towards the bulk diamond, additionally hindering fluorescence detection \cite{Lukosz1977,Hausmann2010}. Consequently, coupling to nanophotonic structures is mandatory to enhance the photon rates from individual color center sensors in single-crystal diamond. Alternatively, attaching nanodiamonds (NDs) containing NV centers to a scanning probe tip provides a scannable NV platform (e.g.,\ \cite{Tetienne2015, Tisler2013}). However, NDs show non-ideal material properties, typically due to excess nitrogen or crystal strain resulting from milling of the material \cite{Santori2010}. Thus, incorporated~color centers may suffer from short coherence time, strong inhomogeneous line spreads and reduced stability~\cite{Bradac2010}. In this review, we thus focus on single-crystal diamond-based sensing techniques. 
Scanning probe sensing demands nanophotonic structures with several properties:
\begin{itemize}[leftmargin=*,labelsep=4mm]
\item	Broadband operation: Section \ref{sec:Introduction} and Figure\ \ref{fig:NVbasics}c illustrate the 100 nm-broad room temperature emission spectrum of NV centers. Room temperature spin read-out under non-resonant excitation is most efficient if the integral emission is detected. Thus, the photonic structure should allow for efficient fluorescence detection from the complete NV emission spectrum. 
\item	Tip-like geometry and suitability for NV centers close to surfaces: a tip-like geometry ensures close proximity of the scannable NV center to the sample even in the presence of alignment uncertainties. The photonic structure needs to be functional for NV centers very close to diamond surfaces. 
\item	Attachment to the feedback system: the nanostructure has to be attached to a force sensor, i.e.,\ a~tuning fork or a cantilever. A miniaturized structure is needed, thus avoiding strong shifting of the resonance frequency due to the additional mass or severe damping of the oscillating system. 
\end{itemize}

Considering the first criterion, resonant photonic structures, like, e.g., photonic crystal cavities~\cite{Faraon2012}, which selectively enhance a specific transition within the emission spectrum, are not fully suitable for sensing applications under ambient conditions. For quantum information, in contrast, these systems are promising as they potentially form a coherent interface between spins and photons in quantum networks \cite{Li2015}. In our case, waveguide-like structures that channel the NV's broad emission into certain spatial modes and direct the light towards the collection optics (nanopillars; see Figure\ \ref{fig:tirphotonicstructures}b) seem more suitable mainly due to their broadband operation. Certain waveguide-based structures, i.e.,\ dielectric antennas \cite{Lee2011}, have been found to enable potentially near unity collection efficiency for NV fluorescence \cite{Riedel2014}. However, they consist of planar multilayer structures and are thus not optimized for scanning probe sensing. In contrast, they might be very suitable for approaches where a diamond chip is used for sensing (wide field approaches, lower spatial resolution, e.g., \cite{LeSage2013}). Taking into account the first two criteria, also solid immersion lenses, where color centers are buried inside diamond half spheres, are not suitable \cite{Marseglia2011}. In this review, we focus on the nanostructure types suitable for scanning probe sensing. For more complete recent reviews of diamond nanophotonics in the context of quantum information, see\ \cite{Hausmann2012a,Beha2012a,Aharonovich2014a,Schroder2016}. 

\begin{figure}[H]
\centering
\includegraphics[width=\textwidth]{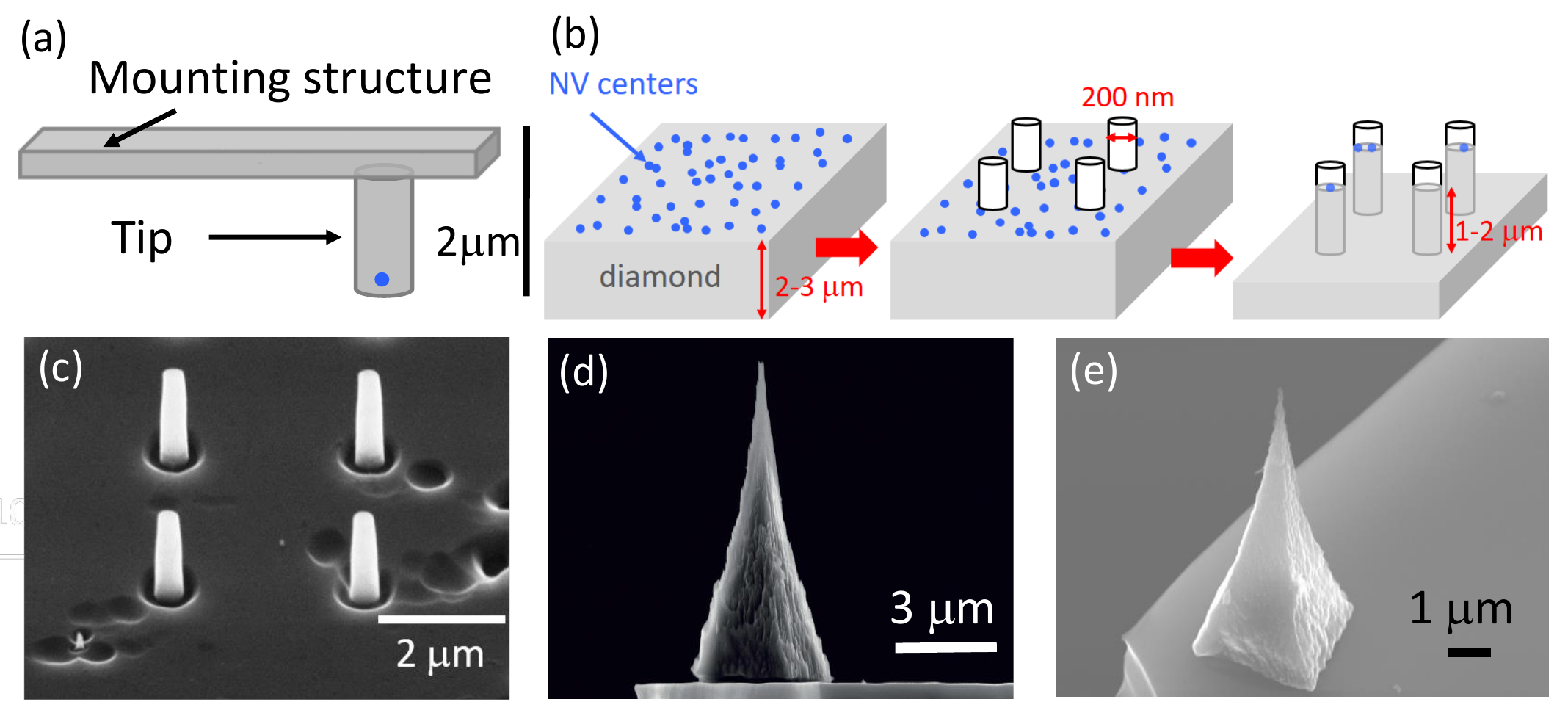}
\caption{\label{fig:scanning_probes} (\textbf{a}) Schematics of an all diamond scanning probe nanostructure, consisting of a cylindrical diamond tip on a thin mounting structure. (\textbf{b}) Illustration of the fabrication process of diamond pillars for sensing: shallowly-buried NV centers are created (blue dots: NV centers; grey: diamond). An~etch mask (white) is created (typically via electron beam lithograph in Hydrogen silsesquioxane
 resist (here Dow Corning, Fox16)). The pillars are etched in a highly-anisotropic etch. NV centers not protected by the etch mask are removed with the surrounding diamond. If the optimal density is chosen, roughly one out of three pillars contains a single NV center~\cite{Appel2016}. (\textbf{c}) Electron microscopy image of diamond nanopillars etched into CVD diamond. (\textbf{d},\textbf{e}) Diamond pyramids formed during a specialized CVD growth process and mounted to silicon cantilevers for scanning probe applications \cite{Nelz2016}. }
\end{figure}

In essence, photonic nanostructures suitable for scanning probe sensing consist of a roughly tip-shaped nanophotonic structure on a thin (typically < 1 $\upmu$m) diamond mounting structure, as shown in Figure\ \ref{fig:scanning_probes}a. First, operational sensors have been presented in 2012 \cite{Maletinsky2012}. The work in \cite{Appel2016, Kleinlein2016} presents more recent nanofabrication approaches. In a further very recent approach, mounting structures holding an array of pillars are used. Consequently, several NV centers are usable, and vector magnetometry might be feasible \cite{Pelliccione2016}. However, this approach introduces the risk that a pillar that is not in direct contact with the sample is used and, thus, an unwanted stand-off distance strongly reduces the resolution.
We start by addressing how large-area, thin, single-crystal diamond membranes can be formed. High-purity, CVD diamond with < 5 ppb substitutional nitrogen and an intrinsically very low density of NV centers is commercially available (see Section \ref{subsec:NVcreation}). Starting from this material, thin, single-crystal diamond membranes can be formed by thinning down diamond plates obtained by polishing (typically several tens of $\mu$m thick). To thin the plates down, reactive-ion etching with different reactive gases in the plasma has been used among them chlorine and oxygen-based recipes \cite{Maletinsky2012, Appel2016}, fully-chlorine-based chemistry \cite{Tao2014}, oxygen/fluorine-based chemistry \cite{Kleinlein2016, Jung2016} and fully-fluorine-based chemistry \cite{Momenzadeh2016}. For all of these approaches except in\ \cite{Kleinlein2016}, argon is added to the plasma to introduce physical etching (sputter etching) to the process. Alternatively, membranes can be created by damaging a buried layer of diamond via ion implantation. When this layer is graphitized upon annealing, it can be chemically etched, and membranes are lifted-off. However, due to strong ion damage, this material is not directly usable and has to be overgrown with a pristine layer of CVD diamond, adding technological complexity to the process \cite{Piracha2016a}. We note that it is not possible to obtain thin, low-stress, single-crystal membranes directly via growth on a non-diamond substrate that might allow for wet-chemical etching of the substrate: for all diamond heteroepitaxy, the starting phase of the growth is highly defective, and forming high-quality, thin membranes includes removing this initial growth areas \cite{Riedrichmoeller2011}. Reliable, large-area fabricating of single-crystal membranes is essential for high-yield fabrication of scanning probe devices and potential future up-scaling of the technology.

The nanostructure forming the tip is either almost cylindrical (termed nanopillar~\mbox{\cite{Babinec2010,Hausmann2010,Hausmann2011,Neu2014,Appel2016})}, pyramidal \cite{Nelz2016} or features a truncated cone shape with a taper angle \cite{Momenzadeh2015,Hausmann2010}. The work in\ \cite{Momenzadeh2015} demonstrated that tapering the pillar enhances fluorescence rates from single NVs. Lithographically-defined etch masks are used to create the pillars, as illustrated in Figure\ \ref{fig:scanning_probes}b. These masks undergo faceting and erosion \cite{Jiang_2016}, thus limiting the pillar length and influencing its shape. For an electron microscopy image of diamond nanopillars, see Figure\ \ref{fig:scanning_probes}c. Mask~erosion is especially critical, as a high-density plasma (mostly oxygen and argon) is needed to enable highly-anisotropic etching of diamond. For the coupling to the nanostructure's photonic modes, not only the placement of the NV center is essential, but also the orientation of the NV's electric dipoles. As indicated in Figure\ \ref{fig:tirphotonicstructures}b, placing the NV's dipoles perpendicularly to the pillar axis is the most advantageous configuration. This situation can be reached in <111>-oriented diamond. Here, NV~centers created during CVD growth preferentially align in the growth direction, with their dipoles in the plane perpendicular to that direction \cite{Lesik2014,Michl2014,Fukui2014}. However, CVD growth in the <111> crystal direction is challenging \cite{Tallaire2014} and still leads to a material with lower quality than the standard <100> growth. In <100> diamond, NV centers align along all four equivalent <111> directions and have an oblique angle of 54.7$^{\circ}$ with the pillar axis. As an alternative, CVD growth in the <113> direction has been investigated \cite{Lesik2015}. In this material, 73\% of NVs form an angle of 29.5$^{\circ}$ with the potential pillar axis and thus bring their dipoles closer to the ideal orientation. The work in\ \cite{Neu2014} demonstrates pillars fabricated into <111>-oriented material. However, so far, optimized crystal orientations have not been used for scanning probe sensing.

To manufacture nanophotonic structures, two fundamentally different approaches exist: first, the structures can be sculpted from bulk diamond material (top-down approach) as discussed above \cite{Babinec2010,Hausmann2010,Neu2014,Appel2016,Jiang_2016}. In contrast, nanostructures might also be created directly during (CVD) synthesis of diamond (bottom up approach \cite{Nelz2016,Aharonovich2013,Zhang2016a}). Pyramidal nanostructures formed via bottom-up approaches (see Figure\ \ref{fig:scanning_probes}d,e) also have advantageous photonic properties \cite{Nelz2016}. However, they~currently show high levels of color center incorporation during growth. This results from growth on non-diamond materials or incorporation of impurities from masks used in the growth. Consequently, these structures are not usable for sensing with single color centers so far. In top-down approaches, plasma-based etching forms the nano-structures. The work in\ \cite{Oliveira2015} shows that a plasma can damage the diamond surface and reduce NV coherence. From this point of view, bottom-up approaches are appealing as they offer the opportunity to avoid plasma etching of diamond. 

As discussed in Section \ref{sec:NVcreation}, shallow NV creation is a crucial step towards functional sensing devices. So far, NVs in nanopillars have been created by ion implantation \cite{Maletinsky2012, Hausmann2011,Appel2016,Momenzadeh2015} followed by structuring the nanopillars via lithography and etching (see Figure\ \ref{fig:scanning_probes}b). Alternatively, NV centers created during growth, but at non-defined depth, were used \cite{Babinec2010,Neu2014}. Thus, the latter pillars are not suitable for scanning probe sensing. A promising alternative approach is spatially localized ion implantation into high-purity nanostructures. Three different approaches have been presented: the~first uses focused nitrogen ion beams for localized implantation \cite{Lesik2013}. In the second approach, a~pierced AFM tip is used as an implantation aperture \cite{Meijer2008}. In the third approach, single ion traps are planned to be used as deterministic sources of implanted ions \cite{Jacob2016}. Also in the context of $\delta$-doping, three-dimensional localization of NV centers has been reported:\ \cite{Ohno2014} uses a mask with nanometric apertures for single NV creation in a $\delta$-doped layer. For the smallest aperture of ($114\pm21$) nm, single~NVs were created in $18\%$ of the apertures. In\ \cite{McLellan2016}, a transmission electron microscope is used for irradiation, achieving sub-micron accuracy in lateral positioning. 

We point out that in parallel to the efforts in nanophotonics, recently electrical read-out of the NV spin state has been proposed alternatively to the challenging optical read-out: NV centers are more likely to undergo multi-photon photo-ionization as long as they are not being shelved to the singlet states. Thus, an NV center in the $m_s=0$ spin state is more likely to be ionized and leads to a higher photo-current. Demonstrations include continuous \cite{Bourgeois2015}, as well as pulsed detection \cite{Hrubesch2017}. Whereas~the read-out contrast can approach 20\%, demonstrations of single NV read-out are still pending. 
 
\section{Recent Advances in NV Sensing \label{sec:advances}}
In this section, we review selected recent advances in sensing with single NV centers. In Section \ref{subsection:nearfield}, we summarize recent steps towards scanning near-field optical microscopy (SNOM) with a single NV center. In Section \ref{subsec:NVadvances}, we present recent results using the magnetic sensing capabilities of NV centers (atomic size sensor, high sensitivity, low-invasiveness) to study physical phenomena at low temperature and/or few atoms level.

\subsection{Near-Field Microscopy with NV Centers \label{subsection:nearfield}}
In recent years, the highly-photostable emission of individual NV centers has triggered several efforts to realize near-field sensing based on NV centers \cite{Tisler2011,Tisler2013,Sekatskii2015,Drezet2015}. In general, near-field-based imaging, where a sample is illuminated by the near-field of a light source, allows a higher resolution than far field-based techniques and is able to beat the Abb$\acute{\textrm{e}}$ limit of resolution.

\begin{figure}
\centering
\includegraphics[width=0.5\textwidth]{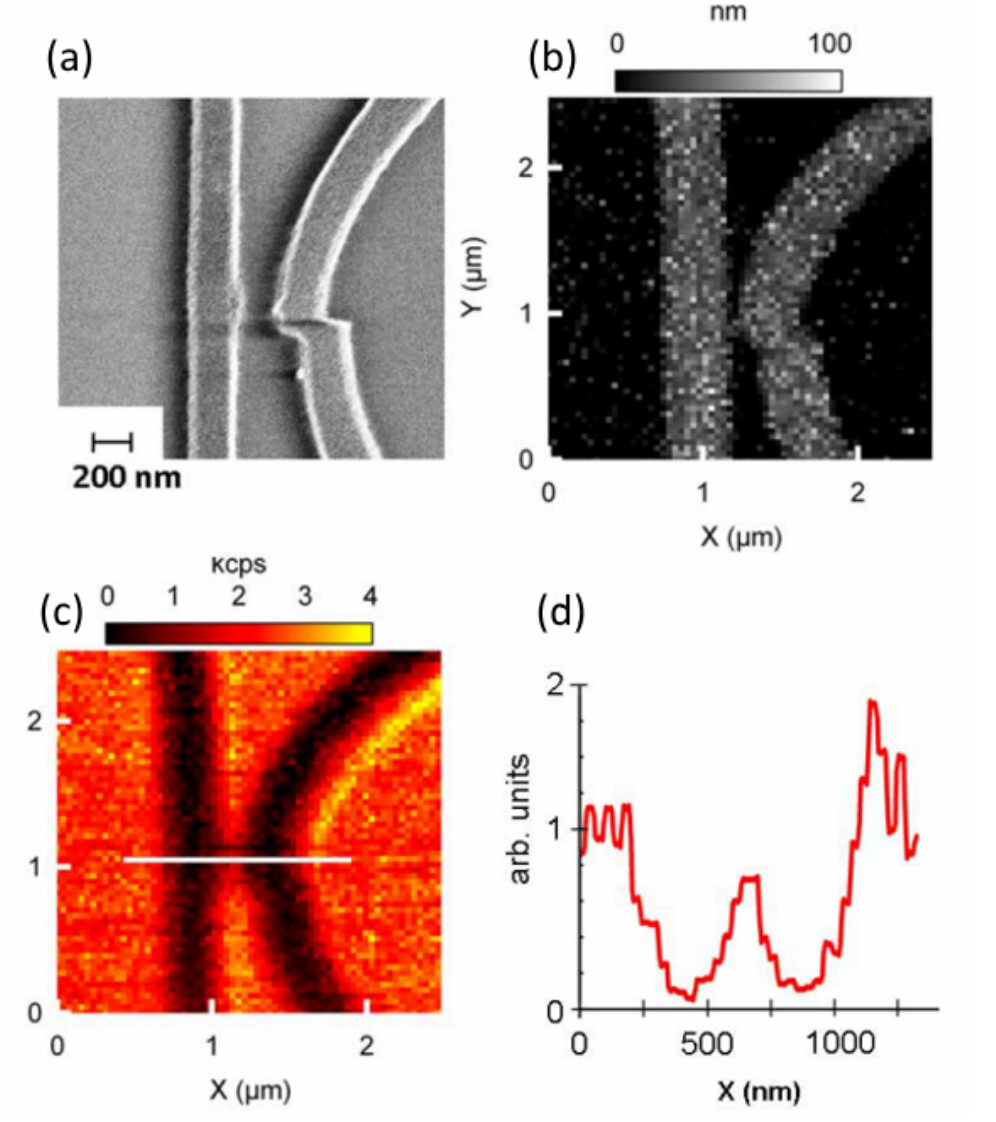}
\caption{(\textbf{a}) Scanning electron micrograph of chromium structures patterned on a fused silica cover slip. (\textbf{b}) Numerically-flattened topography of the same region. (\textbf{c}) Fluorescence scanning near-field optical microscopy (SNOM) image recorded using the single photon tip. (\textbf{d}) cross-cut of the optical intensity along the white line in (c). Reprinted~from\ \cite{Cuche2009a}, copyright the Optical Society of America (2009). }
\label{SNOMImages}
\end{figure}

Single NV centers were used as nano-sized light-source that is brought within < 100 nm of the sample in a SNOM setup \cite{Kuehn2001, Cuche2009a, Drezet2015}. A nanodiamond (ND) attached to the end of a SNOM tip achieved a resolution of around 50 nm when imaging metallic nanostructures, thus beating the Abb$\acute{\textrm{e}}$ limit of resolution (see Figure\ \ref{SNOMImages}). The resolution was limited by the vertical NV-to-sample distance, which is constrained by the ND size and by the fact that the ND was not put in contact with the sample to avoid too strong friction forces at the tip apex.

Förster resonance energy transfer (FRET) was first described as non-radiative energy transfer between a pair of single molecules (dipoles) in close proximity \cite{Forster1948,Dexter1953}. To enable FRET, the emission band of the donor molecule has to overlap with the absorption band of the acceptor molecule, allowing energy transfer from the donor to the acceptor molecule. The transfer efficiency reaches 50\% for a characteristic distance, the so-called Förster distance $\textrm{R}_0$, and drops with the inverse sixth power of the distance. NV centers in a FRET pair with a fluorescent dye have shown $\textrm{R}_0$ = 5.6 nm \cite{Mohan2010a}. Thus, this~effect promises truly nanoscale resolution in near-field-based imaging.

In recent experiments, an ND placed close to the apex of a commercial AFM tip was scanned close to a surface covered with graphene flakes \cite{Tisler2013}. The NV center and the graphene flake form a FRET pair. A Förster distance of 15 nm was found. Sekatskii and co-workers \cite{Sekatskii2015} reported on unsuccessful experiments on FRET transfer between a scanning NV and a dye molecule, despite~previous demonstration of FRET to dye molecules covalently bonded to NDs \cite{Tisler2011}. This finding might relate to a varying quantum efficiency (QE) for NV centers in NDs \cite{Mohtashami2013}, the need for accurate control of the ND surface (graphite layers), as well as to large stand-off distances when attaching an ND to a scanning probe tip. These issues may be addressable in the future using scanning probe devices sculpted out of single-crystal diamond \cite{Appel2016,Maletinsky2012}. So far, no experiments on near-field sensing based on single-crystal scanning probes have been reported.

Finally, we mention that super-resolution images of NV centers in bulk diamond were recorded using stimulated emission depletion microscopy \cite{Rittweger2009}, ground state depletion microscopy \cite{Han2010}, stochastic optical reconstruction microscopy \cite{Pfender2014} and by sampling second- and third- order photon correlation function \cite{Monticone2014}. These techniques beat the Abb$\acute{\textrm{e}}$ resolution limit without the need for near-field imaging.

\subsection{Recent Applications of Single NV Sensing \label{subsec:NVadvances}}

The implementation of scanning NV-based microscopy at cryogenic temperatures forms a recent milestone: in a preliminary realization, cryogenic operation was achieved with a single NV center in bulk material and a scanning magnetic tip \cite{Schaefer-Nolte2014}. More recently, versatile approaches with NV centers in single crystal scanning probes have been presented \cite{Pelliccione2016, Thiel2016}. This technology significantly broadens the range of samples for which NV magnetometry can be applied: solid materials present unique phenomena that only occur at cryogenic temperatures such as superconductivity. An ideal superconductor is supposed to expel all magnetic fields (Meissner effect). However, real~superconductors present points at which magnetic fields penetrate and magnetic vortices form. NV nano-magnetometry was applied for a high-resolution, low-invasive imaging of superconducting vortices (6 K, \cite{Pelliccione2016}, 4.2 K \cite{Thiel2016}). To achieve cryogenic operation, the NV microscope is enclosed in a liquid ${}^4$He cryostat with optical access. Particular care was taken to avoid heating by microwave (MW) and laser excitation. Superconducting vortices were imaged with a resolution below 100 nm with magnetic field sensitivity of 30 $\mu$T$\cdot$Hz$^{1/2}$ \cite{Pelliccione2016} and 11.9 $\mu$T$\cdot$Hz$^{1/2}$ \cite{Thiel2016}, allowing in~\cite{Thiel2016} to verify vortex models beyond the monopole approximation.

Very recently, NV microscopy imaged nanoscale ferrimagnetic domains in antiferromagnetic random access memories \cite{Kosub2017}. We underline that part of these measurements were performed in zero field cooling (ZFC), taking advantage of low-invasive sensing using NV centers.

As discussed above, NV microscopy is often employed to sense static magnetic fields with nanoscale spatial resolution. However, NV centers are also sensitive to magnetic fields oscillating at GHz frequencies: single-crystal NV scanning probes imaged magnetic fields generated by MW currents with nanoscale resolution and sensitivity of a few nA$\cdot$Hz$^{1/2}$ \cite{Appel2015}. The basic idea is to tune an NV spin resonance to the investigated MW frequency using a static magnetic field. The induced coherent spin oscillation frequency (Rabi frequency) then maps the magnetic field amplitude.

Single NV centers also form atomic-sized probes for nuclear magnetic resonance (NMR) signals~\cite{Wrachtrup2016, Mamin2013, Staudacher2013}. The NV spin state is sensitive to the stochastic transverse magnetization at the Larmor frequency of the investigated ensemble of nuclei. Using suitable pulse sequences, it is possible to filter NMR signals and recover NMR spectra. Using NV probes, NMR investigations of small ensembles of nuclei are feasible. In contrast, conventional NMR spectroscopy is limited to macroscopic, thermally-polarized ensembles in high fields. Recent work presents NMR studies of a single protein~\cite{Lovchinsky2016} and of atomically-thin hexagonal boron nitride (h-BN) layers \cite{Lovchinsky2017}. $^2$H and $^{13}$C NMR spectra of ubiquitin proteins were measured \cite{Lovchinsky2016}, paving the way for experimental studies of biological systems at the single-molecule level. Conventional nuclear quadrupole resonance (NQR) \cite{Das1958} spectroscopy yields important information on chemical properties of macroscopic samples. However, it suffers from poor sensitivity due to low thermal polarization. This issue can be overcome using single NV centers that are sensitive to stochastic nuclear magnetization. The stochastic polarization is proportional to $\sqrt{n}$, where $n$ is the number of nuclei, but does not depend on the applied field. Single NV NQR was applied to study thin h-BN flakes, verifying a correlation between the number of atomic layers and a shift in NQR spectra \cite{Lovchinsky2017}.

NMR imaging has already been partially combined with scanning NV microscopy, in the sense that samples on a tip were scanned above a stationary NV. This configuration has been used for magnetic resonance imaging of $^1$H nuclei with a spatial resolution of 12 nm \cite{Rugar2015}. $^{19}$F nuclei were imaged in a calibration grating \cite{Haeberle2015}; widths of $(29 \pm 2)$ nm were resolved \cite{Haeberle2015}. To the best of our knowledge, to date, there are no reports on the use of single-crystal diamond scanning probes for NMR imaging. However, this fact does not represent a general limitation. It most probably relates to the fact that because of challenging nanofabrication procedures, only a few research groups have access to single-crystal scanning probes so far.

\section{Conclusions}
We have summarized recent achievements and the basics of NV center-based sensing. The fabrication of suitable photonic nanostructures that allow scanning probe sensing and high photon rates from single centers, as well as the creation of NV color centers close to the diamond surface are still challenging subjects of research. To date, many recent advances like the creation of NV centers via $\delta$-doping methods, optimized surface treatments and doping of diamond, as well as diamond crystals with optimized orientation have not yet been used in the challenging fabrication of single-crystal diamond scanning probes. Thus, enhanced sensitivities and resolution in NV-based imaging are feasible in the future, given that the recent advances in material sciences are fully transferred to the sensor fabrication. As discussed in this review, many approaches to optimal, shallow NV centers are investigated. The~results considering creation yield and NV properties are not always fully consistent, illustrating the complexity of the field. Moreover, the field could tremendously profit from simplified procedures to upscale the fabrication of optimized sensing nanostructures. 
\vspace{6pt}

\acknowledgments{The authors acknowledge funding via a NanoMatFutur grant of the German Ministry of Education and Research.  Elke Neu acknowledges funding via a PostDoc fellowship of the Daimler and Benz~foundation.}

\authorcontributions{Ettore Bernardi and Elke Neu wrote the manuscript. Selda Sonusen and  Richard Nelz performed the experiments and contributed experimental data (NV spectroscopy, ODMR, nanowires). All authors discussed and commented on the~manuscript.}

\conflictsofinterest{The authors declare no conflict of interest.}

 \bibliographystyle{mdpi}
\renewcommand\bibname{References}

\end{document}